\documentclass[12pt,english]{article}
\usepackage{geometry} 
\usepackage{natbib}

\usepackage{soul}
\usepackage[latin1]{inputenc}
\usepackage{amsmath, amsthm, amssymb, calc, dsfont}   
\usepackage{enumerate}
\usepackage{caption, graphicx} 
\usepackage{authblk} 
\usepackage{setspace, lineno} 
\usepackage{fancyhdr}

\usepackage{tikz}
\usetikzlibrary{shapes,arrows,chains,automata,positioning}
\usetikzlibrary{calc, decorations.pathmorphing}
	\newcommand{\tab}{\hspace*{0.5in}}
	
	\newcommand{\bi}{\begin{itemize} \vspace*{-0.1in}}
	\newcommand{\ei}{\end{itemize}}
	
	\newcommand{\be}{\begin{equation}}
	\newcommand{\ee}{\end{equation}} 
	\newcommand{\ba}{\begin{equation} \begin{aligned}}
	\newcommand{\ea}{\end{aligned} \end{equation}}

\title{H\MakeLowercase{uman judgment vs. theoretical models for the management of ecological resources}}
\author{Matthew H. Holden$^{1,2}$, Stephen P. Ellner$^{1,3}$}
\date{}
\begin{document}

\begin{spacing}{1.5}

This is the pre-peer reviewed version of the following article:\\\\
Holden, M. H., and Ellner, S. P. (2016). Human judgment vs. quantitative models for the management of ecological resources. Ecological Applications. DOI:10.1890/15-1295.\\\\
which has been published in final form at\\\\
	http://onlinelibrary.wiley.com/doi/10.1890/15-1295/full\\\\
For the most up to date version which includes results on how well the human subjects learned and improved their harvest strategy as the game advanced, a discussion that better reflects the authors' current views about the results presented in this study, and many minor corrections, please see the article at the above link or email the corresponding author, Matthew Holden, m.holden1.@uq.edu.au, to request the post-peer reviewed version of the manuscript.
	
\maketitle

\textbf{1:} Center for Applied Mathematics, Cornell University, Ithaca, NY, 14853. USA.\\
\textbf{2:} ARC Centre of Excellence for Environmental Decisions, University of Queensland, 523 Goddard, St Lucia, 4072. Australia.\\
\textbf{3:} Department of Ecology and Evolutionary Biology, Cornell University, Ithaca, NY, 14853. USA.\\
\textbf{Corresponding Author:} Matthew H Holden, m.holden1@uq.edu.au, ARC Centre of Excellence for Environmental Decisions, University of Queensland, 523 Goddard, St Lucia, 4072. Australia.

\begin{abstract}
Despite major advances in quantitative approaches to natural resource management, there has been resistance to using these tools in the actual practice of managing ecological populations. Given a management problem and a set of ecological and economic assumptions, translated into a model, optimization methods can be used to solve for the most cost effective management actions. However, when the underlying assumptions are not met, such methods can potentially lead to decisions that harm the environment and economy. Managers who develop decisions based on past experience and judgment, without the aid of theoretical models, can potentially learn about the system, constantly modifying their mental models to develop flexible management strategies. However, these strategies are often based on subjective criteria and equally invalid and often unstated assumptions. Therefore it is unclear which approach works best for managing biological populations. In this paper, we explore how well humans, using their experience and judgment, manage simulated fishery populations in an online computer game and compare their management outcomes to the performance of a variety of mathematical models. We consider harvest decisions generated using four different theoretical models: (1) the exact same model used to simulate the population dynamics observed in the game, with the values of all parameters known [as a control], (2) the same model, but with unknown parameter values that must be estimated from observed data during the game, (3) models that are structurally different from those used to generate the game dynamics and (4) a model that ignores age structure. Humans on average perform much worse than the models in cases 1 - 3, but in a small minority of scenarios, models produce worse outcomes than those resulting from humans making decisions based on experience and judgment. When the models ignore age structure, they generated poorly performing management decisions, but still outperformed humans using experience and judgment 66 percent of the time.
\end{abstract}

\textbf{keywords:} adaptive management, bioeconomics, conservation, fisheries management, optimal harvest, natural resource management, expert opinion, ecological modeling

\section*{Introduction}
\tab In the past 50 years, environmental management has benefited from major advances in decision science. Perhaps the most influential concept amongst these advances is adaptive management, the iterative process of modeling, hypothesis testing, optimization, acting, and monitoring to reduce uncertainty and maximize net benefits \citep{holling1978,walters1986}. Government agencies, scientists and theoreticians widely agree that adaptive management is the best way to manage a biological population in cases where the benefit of different actions strongly depends on uncertain ecological processes that can be learned through observing system changes in response to management.  \citep{possingham2001, stankey2005, nichols2006, walters2007, williams2012, game2014}. 

\tab While managers often do practice some components of adaptive management by collecting data and making decisions based on their findings, with the exception of a few large scale management programs in fisheries, waterfowl, forestry and conservation \citep[e.g.][]{sainsbury1988, moore2006, nichols2015, johnson2015}, managers rarely use dynamic modeling and optimization, and instead use their experience, intuition and best judgment as a substitute for formal system analysis \citep{johnson20152}. This is despite the fact that many scientists have proposed management plans based on dynamic optimization methods, for a variety of ecological systems, which in theory offer managers substantial cost savings and improved environmental outcomes \citep[e.g.][]{mccarthy2001, westphal2003, gerber2005, wilson2006, asano2008, martin2011, johnson2011, probert2011, rout2014, hughes2014, helmstedt2014}.

\tab One potential reason for the resistance to using mathematical modeling in management is that it's unclear how much modeling and optimization actually improve management outcomes over expert opinion. This is especially a concern when model-based decisions are calculated using \textit{passive} dynamic optimization \citep{johnson20152}. The defining feature of passive optimization is that the method does not consider the value of information while solving for the optimal action, meaning a manager never sacrifices expected gains, given current information, in order to learn about the system and potentially improve long term benefits. 

\tab When the value of improved system knowledge resulting from each action is incorporated explicitly into the objective, the optimization is referred to as active adaptive management. In other words, while both passive and active adaptive management incorporate learning based on observations during system monitoring, only active adaptive management values the future benefit of knowledge resulting from decisions made in the present. Unfortunately, unless a manager is willing to drastically simplify their description of the management problem \citep[e.g.][]{hauser2008}, active adaptive management is computationally infeasible, and hence passive optimization is the predominant method for solving theoretical management problems \citep{johnson20152}.

\tab Humans can possibly use intuition and past experience to incorporate the benefit of learning into decision making, without the aid of mathematical models. However, such decisions are subjective. Can humans use their flexibility to learn about the system to outperform a model-based, passive adaptive management program? Unfortunately, it is difficult to answer this question because experiments in management are, in general, not repeatable. That is, once a manager makes a decision based on their expertise, it is usually impossible to compare the outcome to how well an alternative decision, aided by a mathematical model, would have done.

\tab In this paper we take a first step towards quantifying the economic benefits of using simple dynamic models and passive optimization methods, rather than human judgment, to manage biological populations, by comparing the outcomes from humans and models managing simulated populations. To do this, students in multiple college classes played an online game where they managed a simulated fishery. The data from each game was saved on a server, and therefore we were able to compare exactly how mathematical models would have played, compared to how the students actually played, for each unique instance of the game.


\section*{Methods}

\subsection*{Experiments}
Students played two online ``games", accessed using a web browser, where they earned ``points" corresponding to the profits from managing a simulated herring and a simulated pacific salmon fishery. Below we describe the experiment for the herring fishery game and then explain how the salmon game was different. 

\tab The students played the game using their laptops during the lecture period of two courses, ``Environmental Conservation" at Cornell University and ``Principles of Biology" at Ithaca College, and at the ``Graduate Student Science Colloquium" at Cornell University. Prior to managing each fishery, the students filled out a multiple choice survey that asked them their major, educational experience, fishing experience, and environmental management experience. See \ref{survey} in the online supplementary information for a copy of the survey.

\tab After the survey, each game showed a page of directions describing the fish stock's population dynamics. In addition, Matthew Holden, the game facilitator, read a standardized script aloud to each class, reiterating the points listed on the page. This included statements about the existence of a fishery carrying capacity, measurement error,  environmental randomness out of the managers control and how their performance would be scored. See the online supplementary information for a copy of the game directions. Before starting the game each student was randomly assigned a $\sigma$ between 0 and 0.25. Students with high $\sigma$ experienced large random variation in stock biomass unrelated to their management actions. Before playing the game the students played an 8 turn practice game. This served three purposes: (1) they developed experience with the fishery (2) we used the data from the practice game to identify students who didn't understand the directions and (3) it provided a set of ``past data" for the models and students to use as a basis for making decisions in the future.

\tab Before the user entered their first harvest decision in the practice game, they were presented with 3 harvest data points, and the resulting biomasses from the deterministic version of the model underlying the simulated population dynamics, to give them some context of the range of harvest values they could potentially enter. We chose to use the deterministic model for this purpose so that all users saw the exact same past data before playing the game. A description of the models used to simulate the biomass data observed during the game is presented in the next section, titled ``Simulated population dynamics."

\tab The game showed the user plots of harvest, estimated remaining biomass in the fishery, and cumulative profit at each time step. See Fig. \ref{GamePlay} in the online supplementary information for a picture of the game display. At the beginning of each turn of the game, the user entered an amount of biomass they wanted to harvest into a textbox, clicked enter, and then the remaining biomass, post-harvest, grew according to the models that governed the simulated fishery, and the result was displayed on the screen numerically. In addition, all plots updated, adding the player's harvest choice to the harvest plot, the resulting biomass to the biomass plot and the new accumulated profit to the total profit plot. 

\tab The user's score was the discounted net profit accumulated over the game, with a discount rate of 0.03 and a constant profit of 10,000 dollars per ton of biomass caught. In addition, the user received a bonus added to their score at the end of the game, which was the discounted profit that would have been generated by harvesting all of the remaining biomass left in the fishery after the game was over. The bonus prevents optimal users from harvesting everything on the last turn. Without adding the bonus, the user's score would be highly sensitive to their last harvest decision. This bonus is explained to the user in the game directions (see \ref{gameDirections} in the online supplementary information for a copy of the directions).
  
\tab After a student completed their last turn, the game displayed their score in addition to a leaderboard, which included the scores and initials of the top players in the class, up to that point in time. The leaderboard provided an external incentive to play well. However, the students did not receive a course grade or monetary incentives based on performance. 

\tab Throughout the game, data were stored locally on the user's computer using browser cookies. Upon exiting the game, these anonymous data was then sent to a server, using PHP \citep[a server-side programing language for web development,][]{welling2003}, and stored in a database. These data included the time the user finished playing the game, an anonymous user ID number, the student's answers to the survey questions, the environmental noise variable $\sigma$, total profit (i.e. ``points") and their time series of harvest decisions, resulting biomasses, realizations of environmental noise and measurement error, and in addition the analogous data from their practice game. By recording the environmental noise and measurement error values, experienced by the user, we were able to compare how any strategy (in our case strategies generated by optimization models) would have performed playing that user's exact instance of the game.

\tab After playing the unstructured herring game, the student was directed via a link to the salmon game. Using cookies, the anonymous user ID number from the herring game was saved and recorded along with a unique user ID number for the salmon game as well. In the salmon game, the fishery population dynamics were age-structured, so the game directions also included information on the salmon's life cycle, which consisted of juvenile (1 year-old) and immature (2 year-old) fish survival and growth and adult fish (3 year-old) reproduction. On each turn of the game, the user entered the biomass of adult and immature fish they chose to harvest in two side-by-side textboxes. Plots of the student's harvest and biomass time series data were the same as for the herring fishery, except now each plot had two curves, one for immature fish and one for adult fish. The user could not observe or harvest juvenile biomass. See Fig. \ref{AgeStructuredGamePlay} for a picture of the game display in the age-structured game. 

\tab The user's score in the age-structured game was similar to the unstructured game, except discounted net profit was summed over both adult and immature harvest, and the bonus was the discounted profit that would have been generated by harvesting all of the remaining adult biomass for three years after the game was over (it takes 3 years for the recruits at the end of the game to return to be harvested as adults).

\tab Another goal of this study was to collaborate with instructors to incorporate the game into their curriculum to facilitate active learning. Therefore, while the students played each version of this game multiple times, for pedagogical reasons, students were only asked to try their hardest to score the most amount of points possible during their first game. After everyone had finished their first game, they were allowed to collaborate and experiment, to facilitate students learning the principles of conservation biology, and therefore we did not include the students' latter turns in the analysis.

\subsection*{Simulated population dynamics}\label{simDyn}
The herring fish game was governed by a simple unstructured, one dimensional model, where the manager chooses to harvest $h_t$ tons of biomass in year $t$, and the resulting biomass in year $t+1$, $B_{t+1}$, is a nonlinear function of the biomass that escaped harvest in year $t$, $R(B_t-h_t)$, times a log-normally distributed random number, $z_t$, with mean one and standard deviation $\sigma$. 

\be\label{dynamics1}
B_{t+1}=z_{t}R(B_t-h_t).
\ee

We choose $R$ to be the Beverton-Holt recruitment function, to exclude the possibility of complicated chaotic and periodic dynamics in the absence of harvest,

\be
R(B)=\frac{b_1B}{1+b_2B},
\ee

where $b_1$ is recruitment per unit biomass at low densities and $b_2$ controls the carrying capacity of the population. 

\tab The student managing the population observes a stock biomass of $m_tB_t$, in year $t$, where $m_t$ is a log-normally distributed random variable with mean one and standard deviation 0.025. The small random variation in $m_t$ represents measurement error in assessing the current fish abundance.

\tab The age-structured fish game is based on the life cycle of Coho Salmon, including three independent cohorts that undergo a three stage life cycle. Juvenile fish live in the river and survive and grow into small fish which swim downstream to the ocean where they mature, and finally they swim up stream to spawn and die. The manager sets a total catch of $h_{2,t}$ for immature fish and $h_{3,t}$ for adult fish. Adult fish harvest occurs prior to recruitment, giving population dynamics 

\ba\label{dynamicsAS}
B_{1,t+1} &=z_{t}R(B_{3,t}-h_{3,t})\\
B_{2,t+1} &=z_{t}a_{21}B_{1,t}\\
B_{3,t+1} &=z_{t}a_{32}(B_{2,t}-h_{2,t}),
\ea  

where $a_{ij}$ is the per unit biomass contribution, from age $j$ biomass that escaped harvest, in year $t$, to age $i$ biomass, in year $t+1$.

\tab We parameterized the two models by starting with rough estimates from the literature and then adjusted the values so that the growth rate of our hypothetical herring (unstructured) and coho salmon (age-structured) populations were equal. The reason for using equal growth rates is that when comparing a user's score from managing the population in the unstructured game to the score from the age-structured game, we wanted to make sure that any observed difference was due to demographic structure and not due to differences in the absolute growth rate.

\tab The average 3 year old coho salmon weighs 8.0 pounds  and the average 2 year old salmon weighs about 3.1 pounds \citep{marr1944}. A typical survival probability for pacific salmon populations is 0.8 in good years and 0.28 in bad years \citep{worden2010}. Hence, we fixed $a_{32}=(8 lbs/3.1 lbs)(0.8 + 0.28)/2 \approx 1.4$. Coho salmon are more productive than Herring at low densities, hence we chose to lower salmon recruitment as much as ``believably" possible so that the growth rate in our salmon and herring fisheries matched. To do this, we assumed the average survival probability of juvenile salmon was equal to the estimate for bad  years. Therefore, with the composite parameter of recruitment at low densities estimated in \citep{worden2010} of 60 juveniles per spawner, we let the product of maximum recruitment and juvenile survival be $b_1a_{21} = (0.28) (60 recruits/spawner)(spawner/8 lbs)(3 lbs/recruit) \approx 6.6$. Because juvenile fish are not harvested or observed, the exact value of $a_{21}$ and $b_1$ are unimportant individually, as they only affect the observed immature biomass through their product, and therefore we arbitrarily let them equal 4.4 and 1.5, respectively, so that their product was 6.6. 

\tab Our salmon parameters imply that at low density, the population will grow by a factor of $b_1a_{21}a_{32} = (6.6)(1.4) = 9.24$ over 3 years. We therefore set herring maximum population growth rate at $b_1 = 2.1$ because $2.1^3 \approx 9.24$. The growth rate reported for herring population dynamics ranges from $1.4-1.8$ \citep{bjorndal1987,nostbakken2003}, so while our herring growth rate is high, it is not unreasonably so. Carrying capacity is arbitrarily set to $5,400$ tons, which determines $b_2$ for both models.

\subsection*{Optimal strategies and statistical analysis} Explicit formulas for the optimal harvest strategy, as a function of the parameters, is well known for the unstructured model, and presented in \citet{Reed1979}. A similar optimal harvest rule for the age-structured model is given by \citet{Holden2015}. In both cases the optimal harvest rule is a fixed escapement strategy, where escapement is the biomass that escapes harvest. In other words, the manager leaves a fixed amount of fish in the ocean and this fixed amount of fish is called the escapement. For the parameters in the game, the optimal escapement is $2049$ tons of fish, in the unstructured game, and $556$ tons of adult fish (and all immature fish, i.e. no immature harvest) in the age-structured game \citep[see case 1 in][]{Holden2015}.

\tab The first goal of the experiments was to compare the performance of users to fitted models playing the exact same instance of the game. As a control, we compared both the fitted models' and users' performance to the net discounted profit generated by the optimal constant escapement rule specified above (i.e. the optimal strategy with the true parameters known). 

\tab For all fitted models, parameters were initially estimated using the data generated from the users' eight turn practice game. In the computer's first turn of the game, it follows the optimal strategy (i.e., harvests the population down to the optimal escapement) assuming these parameter estimates are true. After observing the stock biomass resulting from its previous harvest, it re-estimates the parameters using the previous data along with this new data point. It then harvests using the optimal escapement strategy based on the new parameter estimates, and the process is continued until the game is over. 

\tab The parameter estimation for the unstructured game is performed by minimizing sum of squared errors between the log transformed recruitment data, $\log{[m_{t+1}B_{t+1}]}$, and log transformed predicted recruitment under the model, $\log[R(m_{t}B_{t}-h_t)]$, using the function \texttt{lsqcurvefit}, an implementation of the trust-region-reflective algorithm, in \texttt{MATLAB}  \citep{MATLAB}. For the age-structured game, because juvenile biomass is unobservable, the procedure is the same as above, except predicted recruitment is $a_{21}R(m_{3,t}B_{3,t}-h_{3,t})$ and observed recruitment is $m_{2,t+2}B_{2,t+2}$. The transition between immature and adult biomass is estimated similarly. It should be noted that because the mean of the lognormal measurement error is not exactly zero, the above regression is slightly biased. However, correcting for this small bias did not affect the results presented in this paper. 

\tab We consider fitted models with the same functional form (Beverton-Holt recruitment) as the model underlying the simulated population dynamics, and in addition models that incorrectly specify the functional form (discrete logistic and Ricker recruitment). For the age-structured game we also considered escapement rules based on an unstructured Beverton-Holt recruitment model (as in \eqref{dynamics1}). To estimate the parameters for this model, the computer minimizes the sum squared error between the log transformed aggregate biomass data, $\log [m_{2,t+1}B_{2,t+1}+m_{3,t+1}B_{3,t+1}]$, and the predicted biomass, $\log [R(m_{2,t}B_{2,t}+m_{3,t}B_{3,t} - h_{2,t}-h_{3,t})]$. It then harvests the two age classes in proportion to their respective observed biomasses. 

\tab To test whether the percent of optimal profit achieved by the user was correlated with the answers to the survey questions, the standard deviation of environmental stochasticity, and net profit generated during the practice game, not including the bonus, we fitted a linear regression model, using the function \texttt{lm} in \texttt{R} \citep{R}.

\tab Another goal of the experiments was to analyze what strategies the users were deploying and how well different strategies performed compared to others. We compared the user's behavior to three idealized candidate strategies: constant harvest, proportional harvest and constant escapement. Constant harvest means the user enters the same harvest at every time step (harvest $= \beta$, where $\beta$ is the user's mean harvest). Under a proportional harvest strategy the user harvests a constant proportion of the biomass (harvest $=\beta \cdot $ biomass, where $\beta$ is their harvest proportion). Constant escapement, means the user lets a constant amount of biomass escape harvest (harvest $= 0$ if biomass $\leq \beta$, harvest $=$ biomass $ - \beta$ if biomass $>\beta$, where $\beta$ is the biomass they let escape harvest). After fitting these three models to the harvest vs. biomass data generated by each user during their game play, the users were categorized into the three strategy classes based on which model fit had the lowest sum of squared errors. For the age-structured game we repeated the above analysis on adult harvest, for simplicity (because exclusive adult harvest is optimal), but the results reported below are similar if total harvest is used instead.

\section*{Results}

\subsection*{Unstructured population game}

\begin{figure}[h!]
		\includegraphics[width=\textwidth]{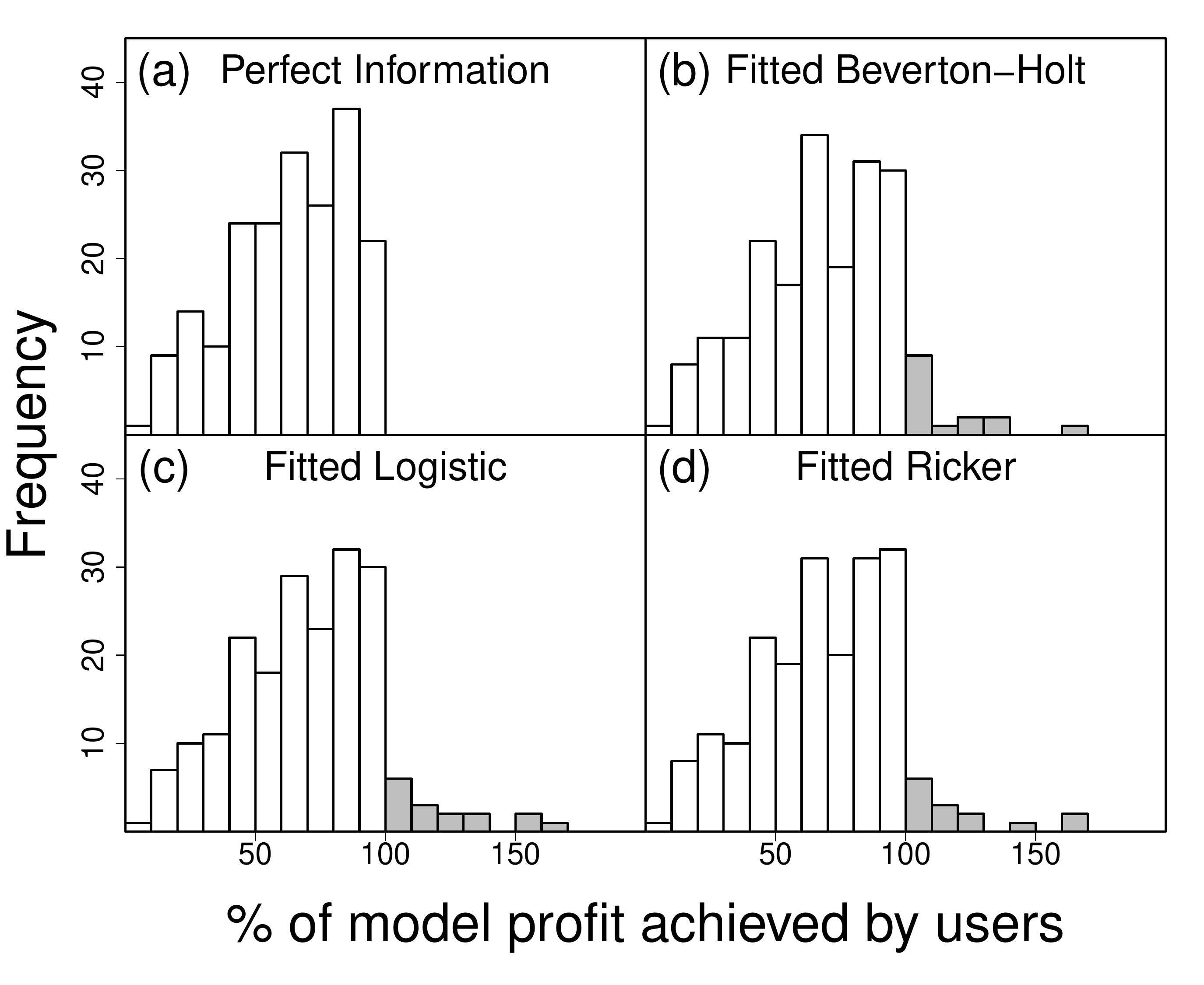}
		\caption{The percent of the mathematical model's total profit achieved by the user, in the unstructured game, when the mathematical model is (a) Beverton-Holt with true parameters, i.e. perfect information (b) Beverton-Holt but with parameters estimated from the data, (c) discrete logistic with parameters estimated from the data and (d) Ricker with parameters estimated from the data. For example, a value of 50 percent means the user generated half the profit the mathematical model did managing the exact same instance of the game. A value of 200 percent means the user generated twice as much profit as the model.}\label{HistUsersVmodels}
	\end{figure}

All human subjects achieved less discounted net profit than would be achieved using the optimal constant escapement strategy with known parameters (Fig. \ref{HistUsersVmodels}a). This was not 100 percent certain to occur, because the optimal strategy is only optimal in expectation, and therefore is not necessarily the most profitable strategy during a run of atypical years. On average, humans scored 65.4 percent of the discounted net profit generated using the optimal constant escapement strategy, and only 11.0 percent of humans achieved over 90 percent of this optimal expected net profit.

\tab Most users performed worse than the escapement rules generated from the models with parameters estimated from the historical harvest data (Fig. \ref{HistUsersVmodels}bcd) even if the model made incorrect assumptions about the underlying recruitment function (Fig. \ref{HistUsersVmodels}cd).

\tab The only significant predictor of the user's performance was their performance in the practice game ($p<0.001$).  Simple linear regression of the user's percent optimal profit on practice game score explained 28 percent of the variation in the user's optimal profit, ($R^2=0.28$, see Fig. \ref{FigPractice}). 

\tab When the practice score was removed as a predictor, the user's level of study (freshman, senior, PhD etc.), academic field of study, and standard deviation of the observed environmental stochasticity, still did not significantly correlate with the user's performance. Two predictors were significant in this model. The five students who responded ``I am considering a career in fisheries management, but have no experience" generated more profit than students that responded ``I am not considering a career in fisheries management" (p = 0.033) and students in Cornell's ``Environmental Conservation" course scored significantly higher than the students in Ithaca College's ``Principles of Biology" course (p = 0.041). However, a linear model with just these two predictor variables only explained four percent of the variation in user performance. It should also be noted that if we group the two students who responded that they actually had fisheries management experience with those five students who indicated a career interest but no experience, the answer to the management experience question does not significantly correlate with the users' scores. This suggests that the sample size for students who were considering careers in fisheries management may be too small to draw any meaningful conclusions.

\begin{figure}[h!]
		\includegraphics[width=\textwidth]{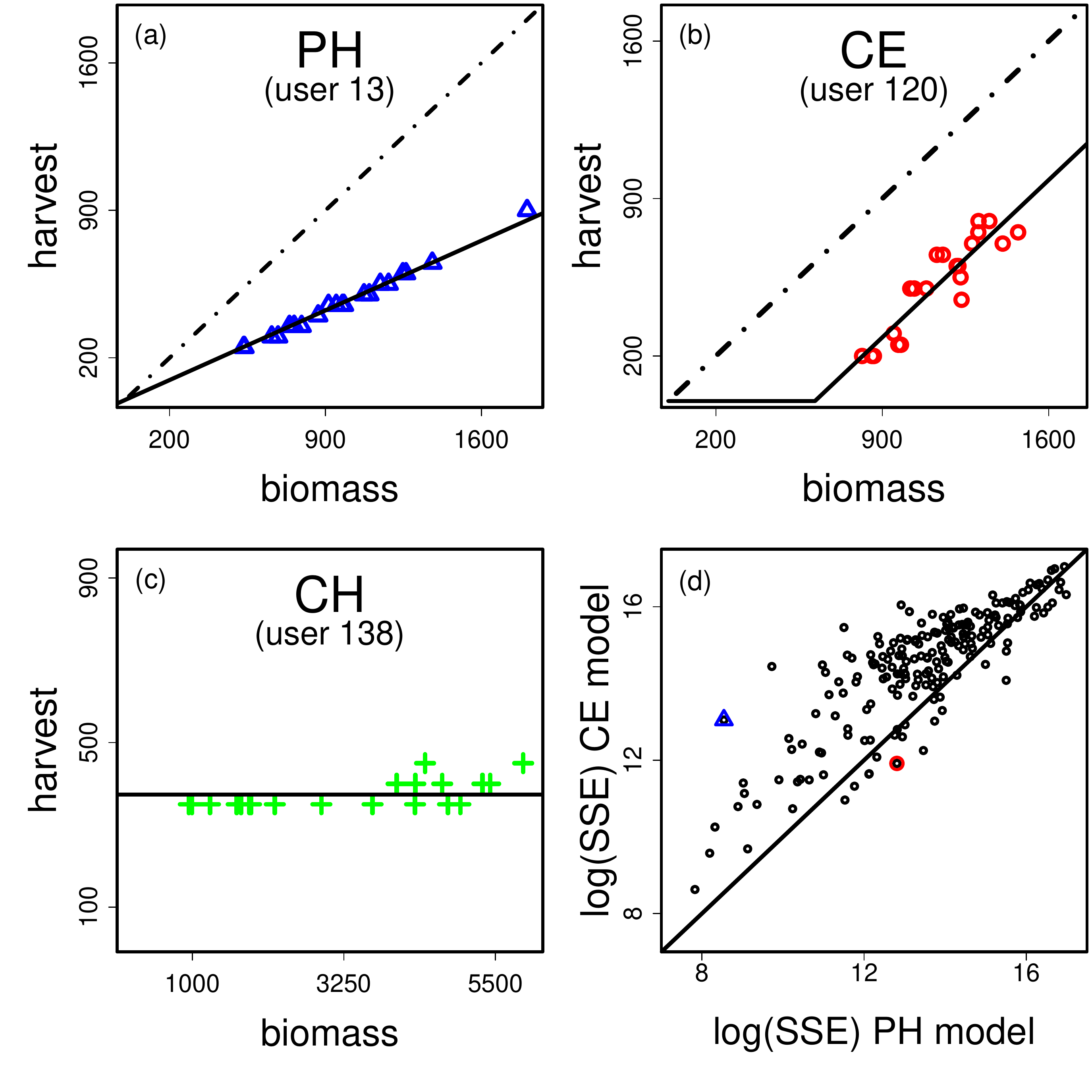}
		\vspace{.1in}
		\caption{ (a-c) An example user's harvest decisions vs. the biomass they observed prior to making those decisions in the unstructured game. The black line is the best model fit, which for user 13 (a) is a proportional harvest strategy (PH), for user 120 (b) is a constant escapement strategy (CE), for user 138 (c) is a constant harvest strategy (CH). (d) The sum of squared error when fitting each user's harvest data, in the unstructured game, to a constant escapement model vs. fitting a proportional harvest model on a log-log scale. Points to the right of the 1:1 line represent users whose variation in harvest is better explained by constant escapement than proportional harvest. The blue triangle and red circle in (d) correspond to the proportional harvester and constant escaper in (a) and (b) respectively.}\label{UserStrats}
	\end{figure}

\tab Classifying the humans' harvest strategies into the three categories: constant harvest, proportional harvest and constant escapement, people harvested a constant proportion (129 users) much more often than allowing a constant escapement (30 users) (Fig. \ref{UserStrats}). Only 39 users were classified as constant harvesters. Many users repeated their harvest decision from the previous turn and the average user only entered 10 unique harvest values over the course of the 21 turn game (Fig. \ref{constHarvesters}). All five users who indicated a career interest in fisheries management were classified as proportional harvesters. Of the two users with actual management experience, one was a constant harvester and the other was a proportional harvester.  The optimal policy type (constant escapement) was the only strategy not used by students with fisheries interests or experience. 

\begin{figure}[h!]
		\includegraphics[width=\textwidth]{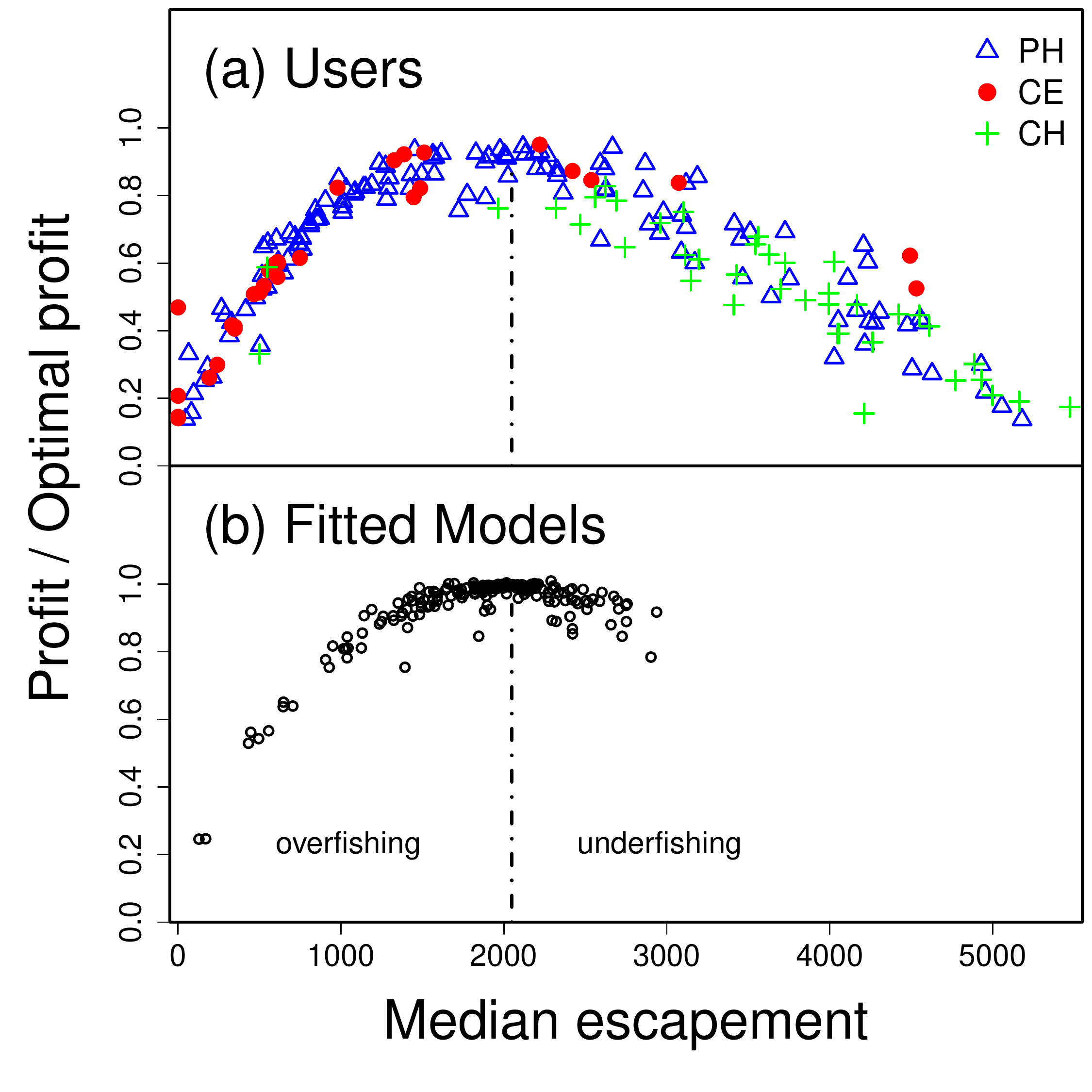}
		\caption{(a) The profit generated by each user, in the unstructured game, relative to the net profit the optimal strategy with perfect information would generate in the corresponding instance of the game, as a function of the median amount of fish the user let escape harvest. The red circles, blue triangles and green pluses are for users who used constant escapement (CE), proportional harvest (PH), and constant harvest (CH) strategies, respectively. (b) the analogous proportion of optimal profit generated by the fitted model vs. the median of escapements chosen by the model after it fit a recruitment function to the data during each turn of the game. The dotted line is optimal escapement under perfect information. } \label{underOverFish}
	\end{figure}

\tab Forty-five percent of humans allowed less fish to escape harvest, on average, than the optimal value (Fig. \ref{underOverFish}a). In other words 45 percent of users over-fished the population while 55 percent of users under-fished the population. If we were to re-classify users whose median escapement was within $q$ percent of the optimal value as neither under nor over-fishing, the result that there are roughly the same number of over and under-fishers holds for all $q<70$. So while many humans harvested sub-optimally, over and under-fishing mistakes were equally likely. 

\tab Students who used constant escapement strategies (circles in Fig. \ref{underOverFish}a) were more likely to over-fish than to under-fish. Proportional harvesters (triangles in Fig. \ref{underOverFish}a) both over and under-fished and constant harvesters (pluses in Fig. \ref{underOverFish}a) were much more likely to under-fish. Note that constant harvesters really can only under-fish, because if they were to over-fish, the biomass would eventually decrease to the point where their constant harvest would crash the fishery, at which point they would have to abandon the constant harvest strategy.

\begin{figure}[h!]
		\includegraphics[width=\textwidth]{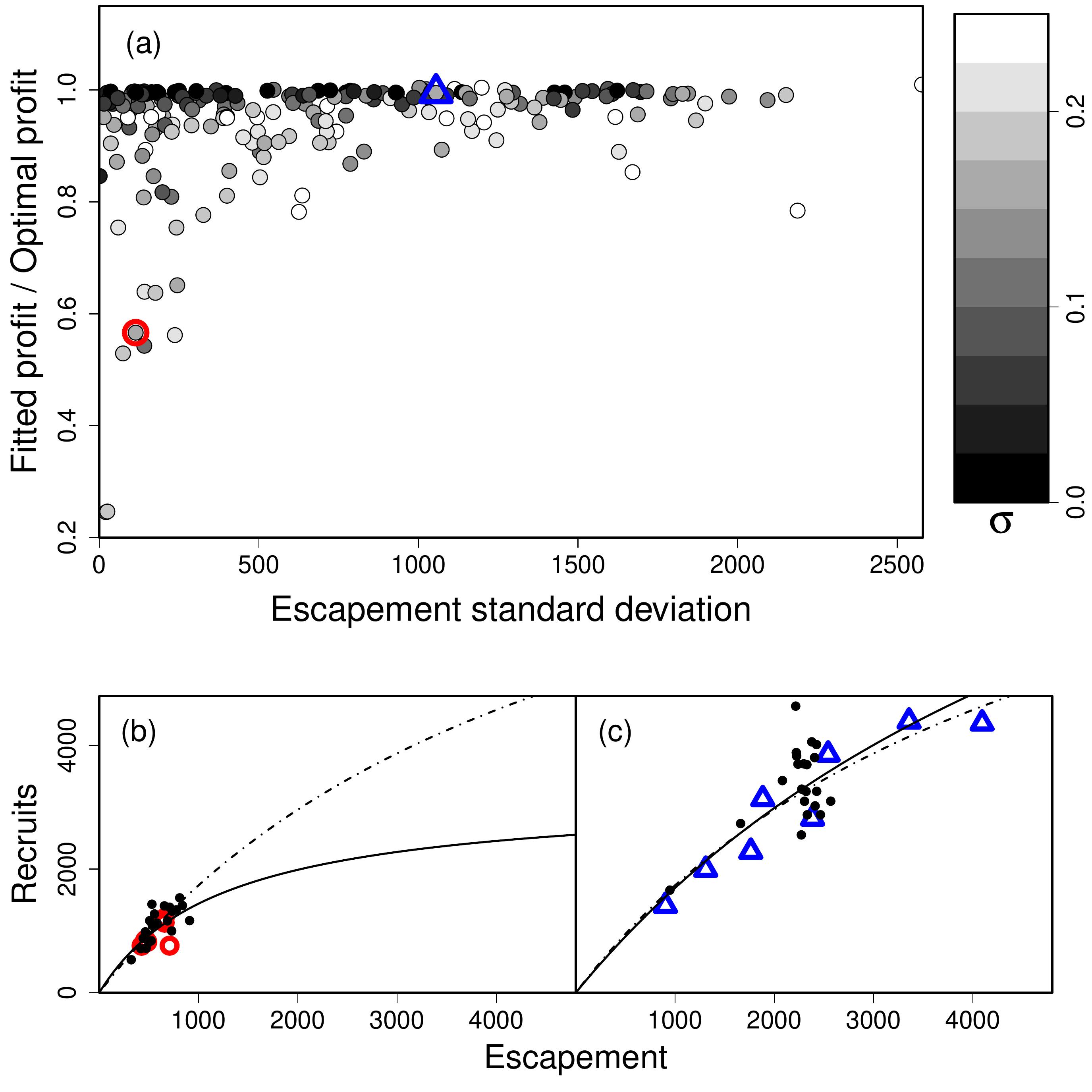}
		\vspace{.1in}
		\caption{(a) The profit generated from the strategies using the fitted Beverton-Holt model relative to the optimal profit under perfect information, as a function of the standard deviation in practice game escapement, generated by the user. Dark and light circles are for instances of the games with low and high levels of environmental stochasticity respectively. (b-c) The true recruitment function (dashed line) and fitted recruitment function (solid line) for two instances of the game, [these examples are highlighted by a red circle and blue triangle in (a)], where the fitted model generates unprofitable escapement strategies (b - red circle) and profitable ones (c - blue triangle). The open symbols are recruitment data generated by the user in the practice game, whereas the smaller filled points are generated by the fitted model when playing the actual game.}\label{sigmaSigEsc}
	\end{figure}

\tab When harvest rules generated by the fitted Beverton-Holt recruitment model performed poorly, this was most often due to over-fishing rather than under-fishing (Fig. \ref{underOverFish}b). Poor model performance was due to a combination of two reasons: (1) during the practice game the user allowed similar amounts of biomass to escape harvest on every turn, generating poor data for model fitting, and (2) the standard deviation of the environmental noise was high  (Fig. \ref{sigmaSigEsc}). When both of these conditions are true, the data can misrepresent the recruitment function (Fig. \ref{sigmaSigEsc}b compared to \ref{sigmaSigEsc}c) and lead to a poor escapement strategy. Despite the poor escapement strategies that sometimes resulted from the fitted models, they still were less frequent and generated more long term discounted profit than the worst users (compare the low points in Fig. \ref{underOverFish}a to \ref{underOverFish}b). 

\tab In the supplementary information we show how quickly the parameter estimates, during the Beverton-Holt model fitting, converged to the true values governing the game dynamics. In general, poor model fits were rather common. For example, when decisions were made based on the Beverton-Holt model with parameters estimated from biomass observations during the game, even after all 21 turns, the estimated value for $b_2$ was off by more than 25 percent over 40 percent of the time (see turn 21 in \ref{parmConv}a). Parameter estimation of $b_1$ was more accurate (\ref{parmConv}b), and for both parameters the poor estimates improved as time moved forward (\ref{parmConv}ab). Still, in many of the cases where parameter estimates were off, the fitted models made more profitable decisions than the users, despite the poor model fit (\ref{HistUsersVmodels}b). 

\clearpage

\subsection*{Age-structured population game}
\begin{figure}[h!]
		\includegraphics[width=\textwidth]{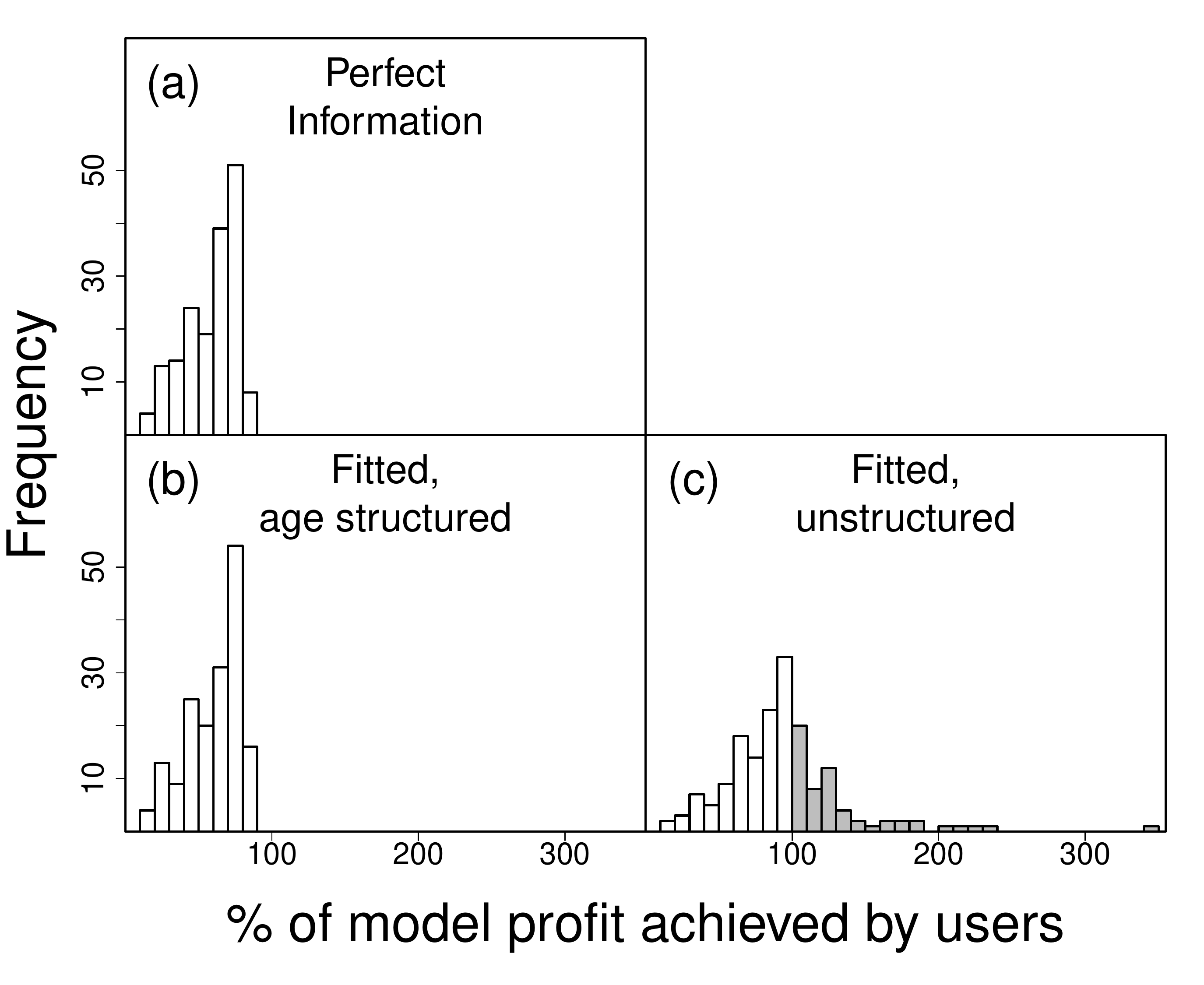}
		\vspace{.1in}
		\caption{The percent of the mathematical model's total profit achieved by the user, in the age-structured game, when the mathematical model is (a) age-structured with true parameters, i.e. perfect information (b) age-structured but with parameters estimated from the data, and (c) unstructured with parameters estimated from the aggregated (immature + adult) biomass data. For example, a value of 50 percent means the user generated half the profit the mathematical model did, managing the exact same instance of the game. A value of 200 percent means the user generated twice as much profit as the model.}\label{histAgeStructure}
	\end{figure}

In the age-structured model, the average user achieved 63.6 percent of the optimal profit achieved by a model with perfect information. The most profitable user scored only 84.3 percent of the optimal profit, in comparison to the best performer in the unstructured population game who scored over 95 percent of optimal profit. On the opposite end of the spectrum the worst users in the unstructured population game only scored 7.2 percent of optimal profit, while in the age-structured game the worst user scored 11.8 percent of optimal profit. A full distribution of the relative performance of the users compared to the optimal policy in the age-structured game is given in Fig. \ref{histAgeStructure}. The improved performance by the worse players, despite the age-structured game being more complex, was due to the fact that this game includes three independent cohorts. Even if one or two cohorts were driven to low levels, some harvest could be achieved in the remaining turns as long as one cohort was still abundant. A player could make one very bad decision, and learn from it, without collapsing the entire fishery.

\begin{figure}[h!]
		\includegraphics[width=\textwidth]{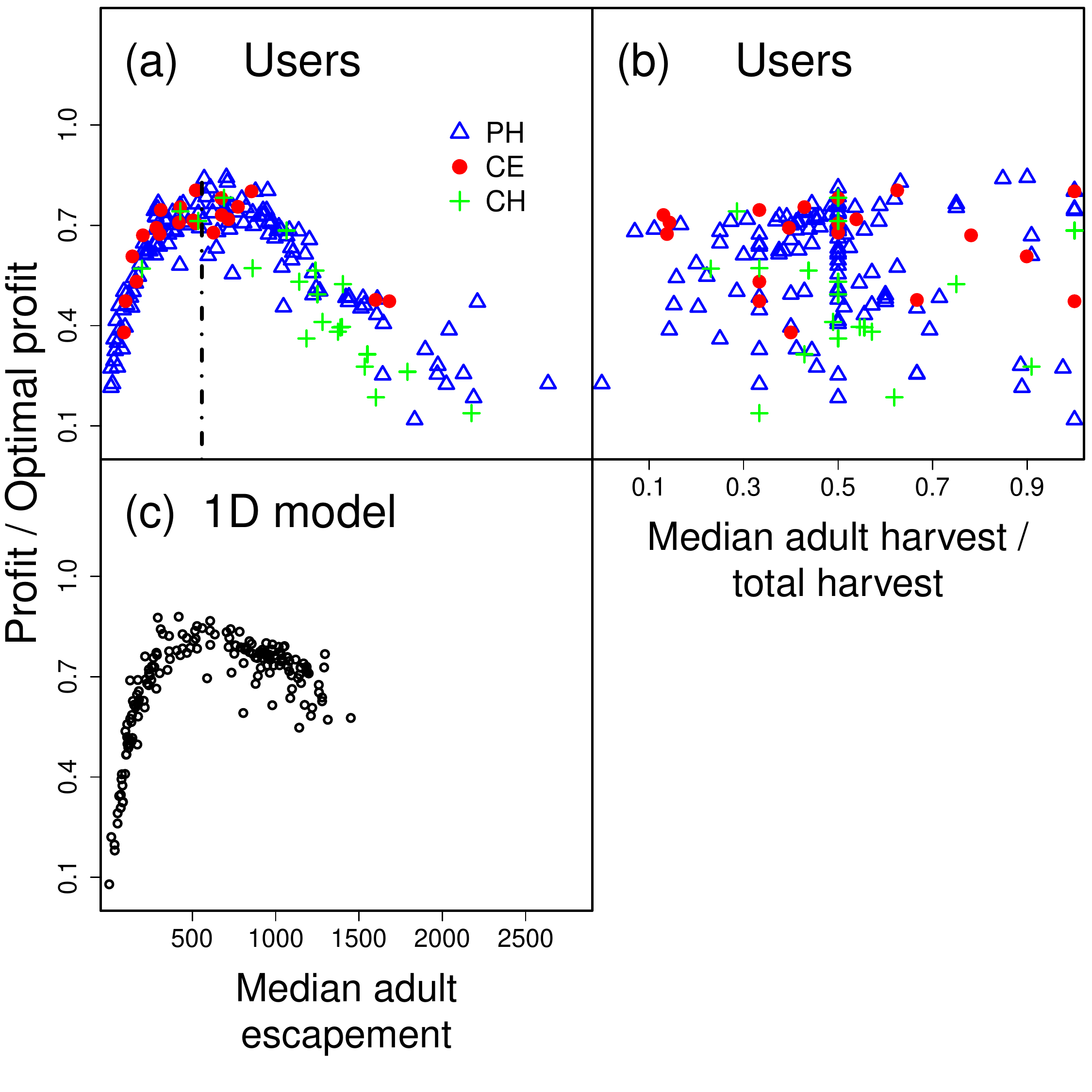}
		\vspace{.1in}
		\caption{(ac) The proportion of optimal profit, in the age-structured game, generated by (a) the user and (c) fitted unstructured model, with parameters estimated from the aggregated (immature + adult) biomass data, for each game as a function of the median escapement chosen. (b) The proportion of optimal profit generated by the user as a function of the median proportion of harvest allocated to adult biomass during the game.}\label{UnderOverFishAge}
	\end{figure}

\tab The user's performance in the age-structured game was mainly determined by their overall fishing pressure and not their decision of which age class to fish (compare Fig. \ref{UnderOverFishAge}a to \ref{UnderOverFishAge}b). The majority of users harvested more immature biomass than adult biomass, despite exclusive adult harvest being the optimal strategy (\ref{UnderOverFishAge}b). Similar to the simple unstructured game, users who deployed a constant escapement strategy (for adults) were more likely to over-fish (\ref{UnderOverFishAge}a).

	\clearpage

\tab The average escapement strategies resulting from the age structure model with parameters estimated from historical data (fitted age structure model) achieved 98.4 percent of the optimal profit, even better than in the unstructured game. Even in the fitted model's lowest performing game, it achieved 78.9 percent of optimal discounted net profit, far better than even the median user. This is for two reasons (1) users in the practice game tended not to let the same amount of adult fish escape harvest every turn, producing good data for model fitting, and (2) the transition rate from immatures to adults was always estimated well, because it is a single parameter that can be estimated independently from recruitment, whereas the recruitment function requires two parameters to be estimated simultaneously. The result of point (2) is that the models always fished from the correct age class.

\tab The average escapement strategy, generated by fitting a one dimensional unstructured population model to the aggregate age-structured data, achieved 72.3 percent of the optimal profit. This represents a 13.5 percent gain in profit over the average human operating solely on intuition. Only 58 users, out of 172, generated more profit than would have been obtained by harvesting based on the fitted unstructured model. However, for instances of the game where the unstructured model generated low discounted net profit, the model's proposed escapement rule crashed the fishery, by letting very little biomass escape harvest. These strategies generated less discounted net profit than the least profitable users (compare the lowest points in Fig. \ref{UnderOverFishAge}a to \ref{UnderOverFishAge}c).  

\section*{Discussion}
Many mathematical tools exist to improve decision making in environmental management, including methods from optimization and optimal control. Yet managers are still resistant to using these tools to develop management plans, and instead rely mainly on their experience and intuitive judgment \citep{johnson20152}. At least one reason for this is that it is often unclear what a manager may gain by using quantitative methods, especially if the dynamics of the managed system are not well understood. 

\tab In this paper we studied optimal escapement strategies for the management of simulated fisheries, developed using simplified models of fish stock dynamics, and tested their performance compared to humans managing the simulated population using only their experience and judgment. The models performed better than the users, on average, even when the models mis-specified recruitment or state variables. However, in the age-structured game, the worst outcomes produced by the simplified unstructured model were worse than the worst outcomes generated by the users.

\tab Each quantitative model-based approach to managing the simulated fishery used a single equation or system of equations with unknown parameters. Alternatively, a manager could develop a set of candidate models representing alternative hypotheses about the system (e.g. an age-structured and unstructured model) and then require quantitative methods resolve structural uncertainty \citep{williams2001,nichols2015}. Our results on using unstructured models to manage age-structured populations are an example of a worse case scenario where the a manager's candidate model set does not contain a model that approximates the system well. Even in such a case, on average, quantitative methods out performed human judgment in our experiment.

\tab Users and fitted models tended to make different types of mistakes. An equal number of users over-fished vs. under-fished the stock. However, when management based on fitted models failed, it was almost always due to overfishing. 

\tab We found that even when the model is perfectly specified, and only needs parameter estimates from the data, it still can perform worse than a human using intuition alone, especially when environmental stochasticity is high and prior management decisions have all been similar. The lack of data with sufficient variability in stock abundance to estimate parameters well is likely common in fisheries management because overexploited populations are ubiquitous, and therefore the time series data of fish stock abundances may often contain only population sizes well below carrying capacity. In such cases, recruitment curves may often be incorrectly estimated and our simple models will naively suggest that it is optimal to keep overfishing. This suggests that passive adaptive management, choosing the best strategy, based on the current knowledge of the system, to optimize some objective, without any regards to the information gained by deploying that action can potentially lead to poor performance even when model structure is correctly specified.  

\tab Our results suggest that probing the system by performing an action that is suboptimal given the manager's current belief about the system, but that will reveal information that improves management in the future might be desirable in such scenarios. Incorporating the economic benefits of learning from experimentation explicitly into the optimal decision problem, known as active adaptive management, has been studied within the context of harvested populations.  However, due to computational limitations solutions are always limited to cases with one of the three following assumptions: (1) both the probability distribution specifying environmental stochasticity and all parameters in the recruitment function are perfectly known, except for a single parameter to be estimated from the data \citep{walters1981,ludwig1982}, (2) there is a small number of candidate models, with all parameters fixed within each model \citep{williams2001}, or (3) only a small number of actions and system states are admissible  \citep[e.g. action = harvest or not, fishery state = robust, vulnerable, or collapsed, ][]{hauser2008}. 

\tab Unfortunately, the problem of choosing an optimal escapement level in our game, using the principles of active adaptive management, is computationally infeasible given current algorithms and computing power because our game allows for an infinite set of possible actions and states, governed by unknown parameters and unknown variability in environmental noise.  

\tab It is rather alarming that even in the most optimistic case, where the underlying dynamic model is known and parameters have to be estimated from the data, passive adaptive management can fail to achieve desirable results. However, the alternative of letting humans manage our simulated fishery based solely on their experience and judgment typically led to much worse outcomes. Because mathematical models usually improved management outcomes in our experiment, we would recommend modeling be more widely adopted in management. However, models should not necessarily be considered as a replacement for manager expertise. Our results show that in some cases human intervention may be required when models appear to recommend risky management decisions. 

\section*{Acknowledgments} This project was funded by the National Science Foundation grant ``Computational Sustainability: Computational Methods for a Sustainable Environment, Economy, and Society" (award number 0832782). We especially thank our web developer Mikhail Yahknis for building the fisheries game website and Jon Conrad and Jan Nyrop for helpful comments on the manuscript.

\clearpage


\section{Supplementary Information}
	\renewcommand{\thefigure}{S\arabic{figure}}
	\setcounter{figure}{0}
	\renewcommand{\thetable}{S\arabic{table}}
	\setcounter{table}{0}
	
	\begin{table}
	\caption{Below is a copy of the survey.}\label{survey}
						How many times have you played this game before?
						\begin{itemize}
						\item possible answers: ``0",``1", ``2",... ``15", ``more than 15", ``I've only played a different version of this game"
						\end{itemize}

						What best describes your education level? 
						\begin{itemize}
							\item possible answers: ``No college", ``Some college or associates degree, but not currently in college", ``Freshman (1st year)", ``Sophomore (2nd year)", ``Junior (3rd year)", ``Senior  (4th year or greater)", ``completed Bachelor's degree", ``In PhD, MS, MA, or MEng program", ``In other post-bachelor program (e.g. JD, MBA, MD, MFA)", ``completed MS, MA, or MEng degree", ``completed PhD degree", ``completed other post-bachelor degree (e.g. JD, MBA, MD, MFA)"
						\end{itemize}

						What best describes the field of study for your highest degree?					
							\begin{itemize}
								\item possible answers: ``No college", ``Math, Statistics, or Computation",`` Ecology, Natural Resources, or Environmental Biology", ``Economics", ``Political Science or Government",`` Environmental Engineering", ``Other Engineering", ``Other Biology", ``Physical Sciences", ``Social Sciences", ``Humanities", ``undecided"
							\end{itemize}

						What best describes your experience fishing? 
							\begin{itemize}
									\item possible answers: ``I have never fished",``I fish or have fished, but less than once per year", ``I fish recreationally, at least once per year, but not for a living", ``I fish or have fished for a living"	
							\end{itemize}
	
						What best describes your experience managing fisheries?
							\begin{itemize}
									\item possible answers: ``I have work/intern experience managing fisheries, more than 10 years", ``I have work/intern experience managing fisheries, between 10 and 2 years", ``I have work/intern experience managing fisheries, less than 2 years", `` I am considering a career in fisheries management, but have no experience", `` I am not considering a career in fisheries management"	
							\end{itemize}
	
						Do you have work experience managing any (non-human) biological population outside of fisheries? 
							\begin{itemize}
									\item possible answers: ``I have work/intern experience managing game (hunted populations)", ``I have work/intern experience in conservation biology", ``I have work/intern experience in forestry", ``I have work/intern experience in agricultural management", ``I have work/intern experience managing other biological populations", ``I am considering careers in managing biological populations, but have no experience", ``I am not considering careers in managing biological populations"
							\end{itemize}
							
	\end{table}
	
	\clearpage

 \begin{figure}[h]
     \includegraphics[width=\textwidth]{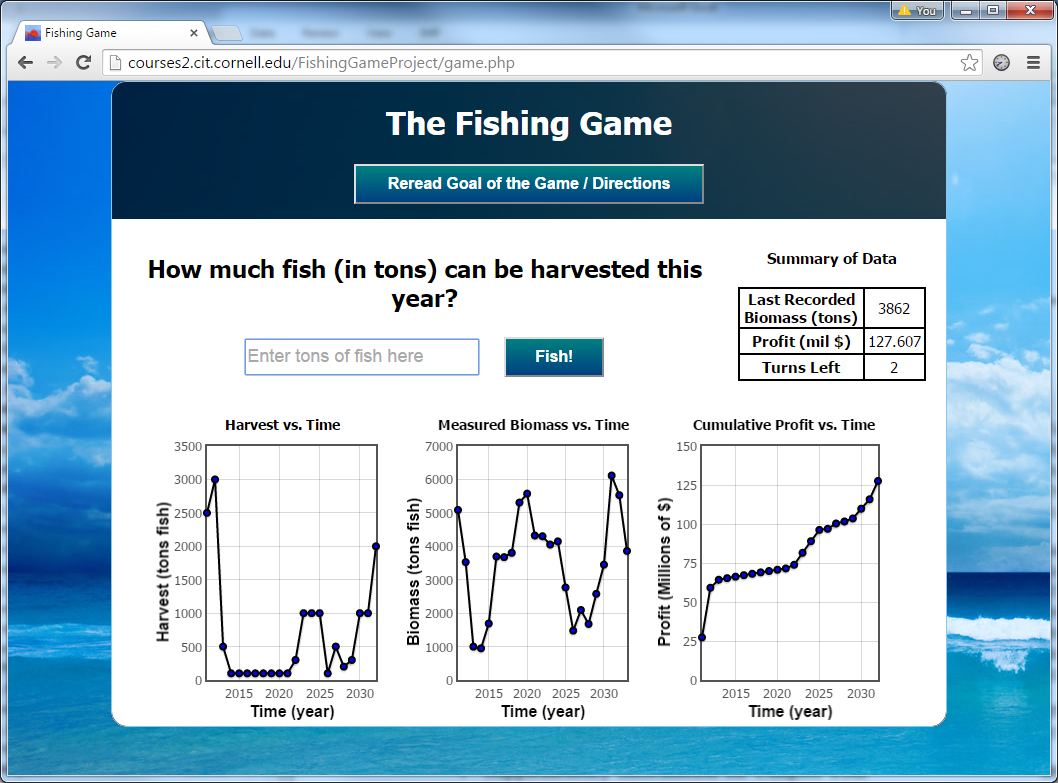}
     \caption{Game play for the unstructured herring fish game.}\label{GamePlay}
 \end{figure}

  \begin{figure}[h]
      \includegraphics[width=\textwidth]{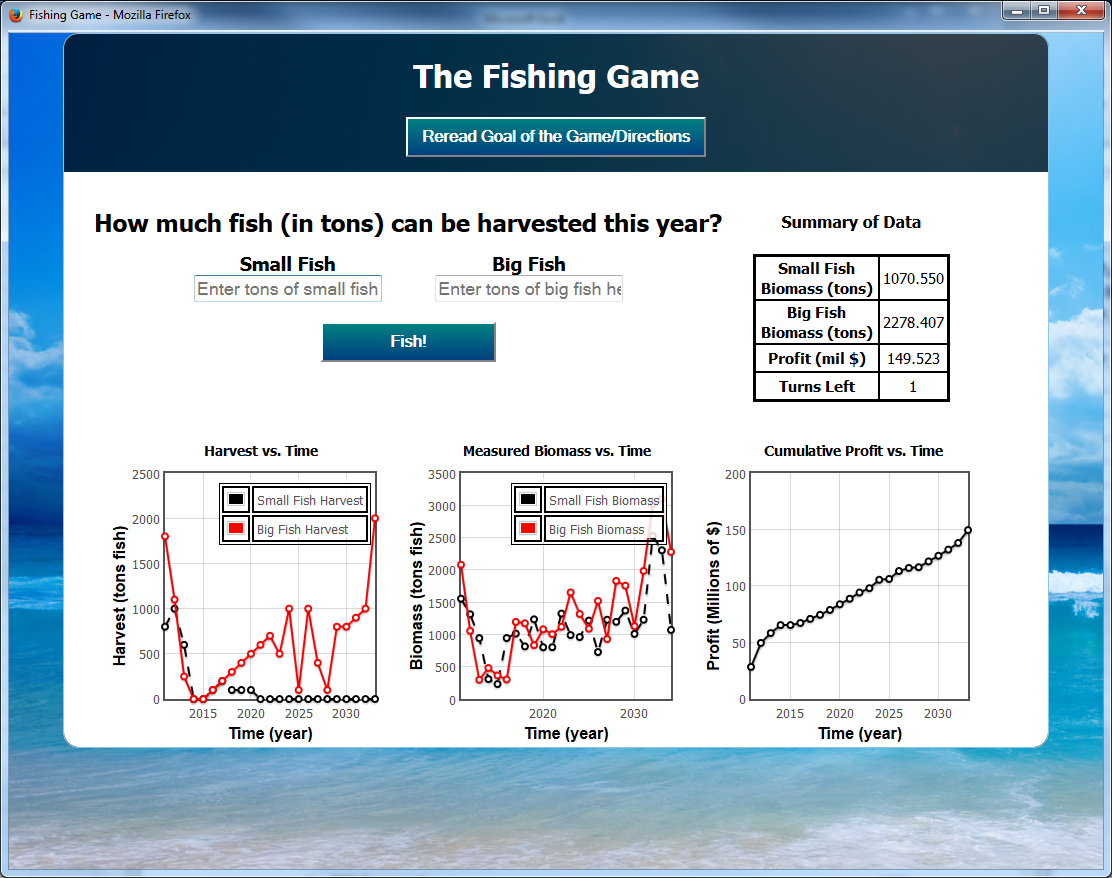}
      \caption{Game play for age-structured salmon fish game.}\label{AgeStructuredGamePlay}
  \end{figure}
  
\clearpage

 \begin{figure}[h]
     \includegraphics[width=\textwidth]{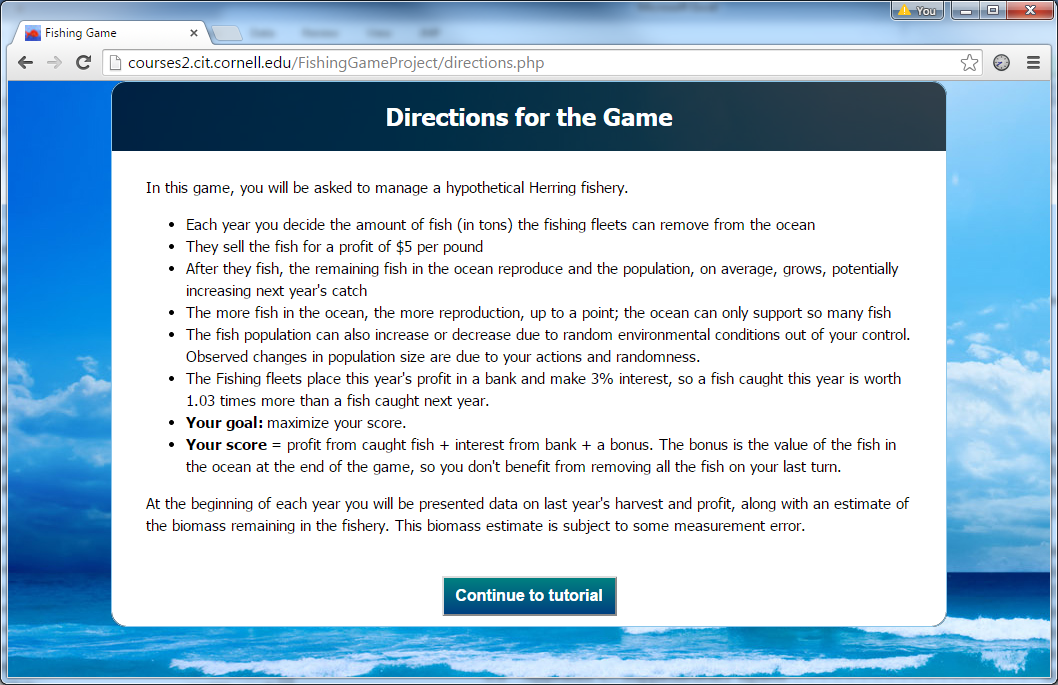}
     \vspace{.1in}
     \caption{Directions for unstructured herring fish game.}\label{gameDirections}
 \end{figure}
 
  \begin{figure}[h]
      \includegraphics[width=\textwidth]{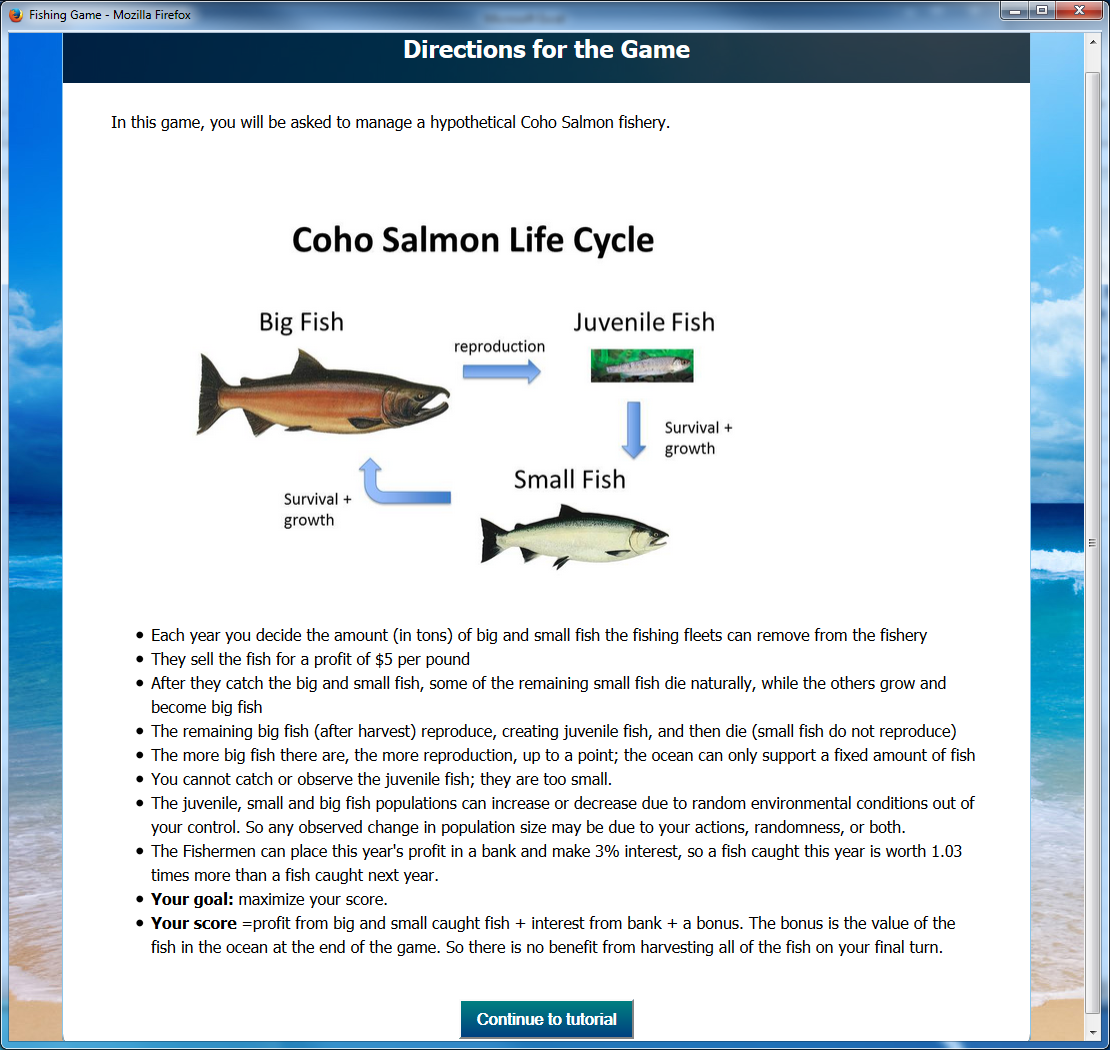}
      \vspace{.1in}
      \caption{Directions for age-structured salmon fish game.}\label{ageGameDirections}
  \end{figure}
  
\clearpage

	\begin{figure}[h!]
		  \includegraphics[width=.98\textwidth]{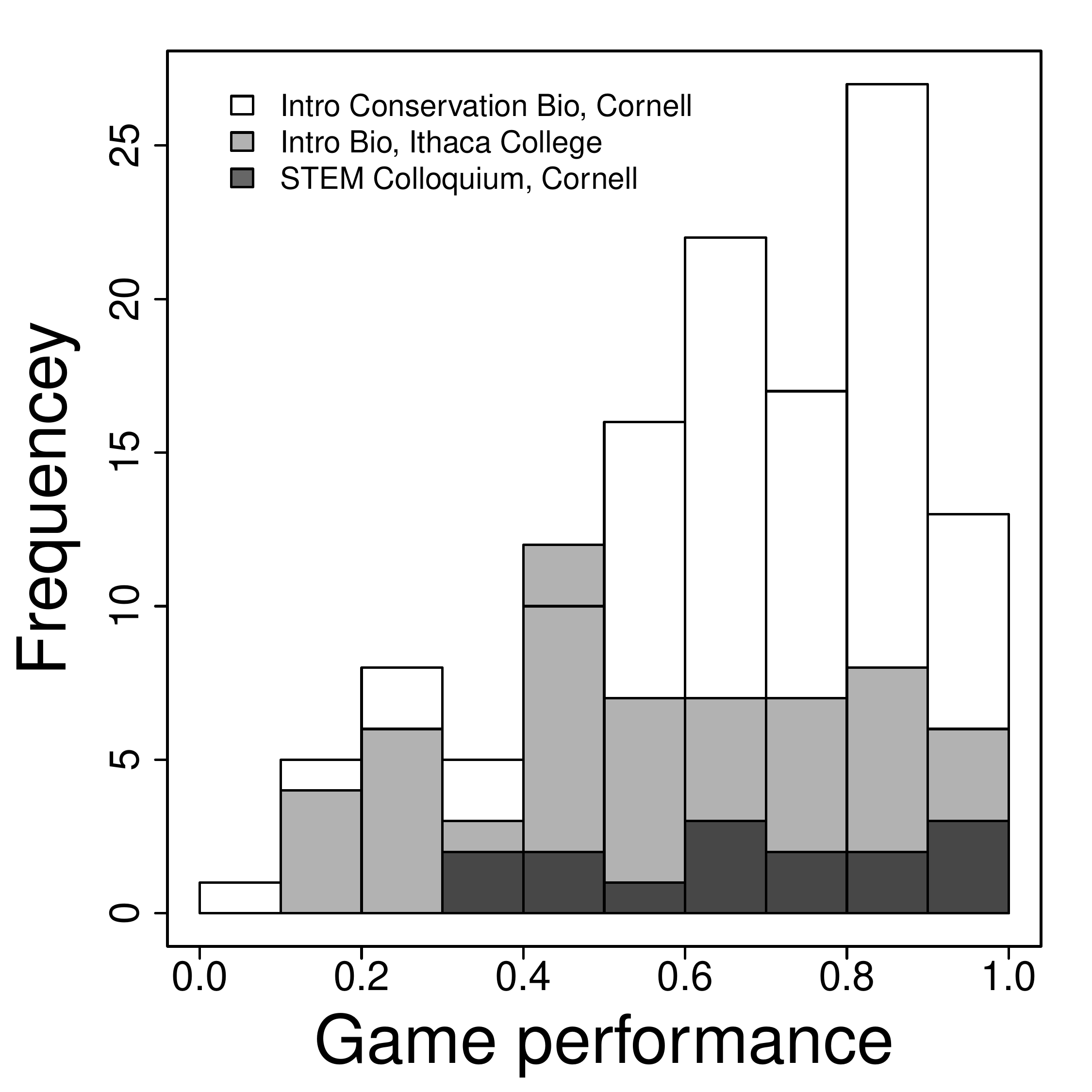}
		  \vspace{.1in}
	 \caption{A histogram of percent of optimal profit achieved by the users for each class. Note the histograms are overlaid rather than stacked.}\label{FigHistSessionPerformance}
	\end{figure}

	\begin{figure}[h!]
          \begin{minipage}{.495\textwidth}
          (a)\\
           \includegraphics[width=\textwidth]{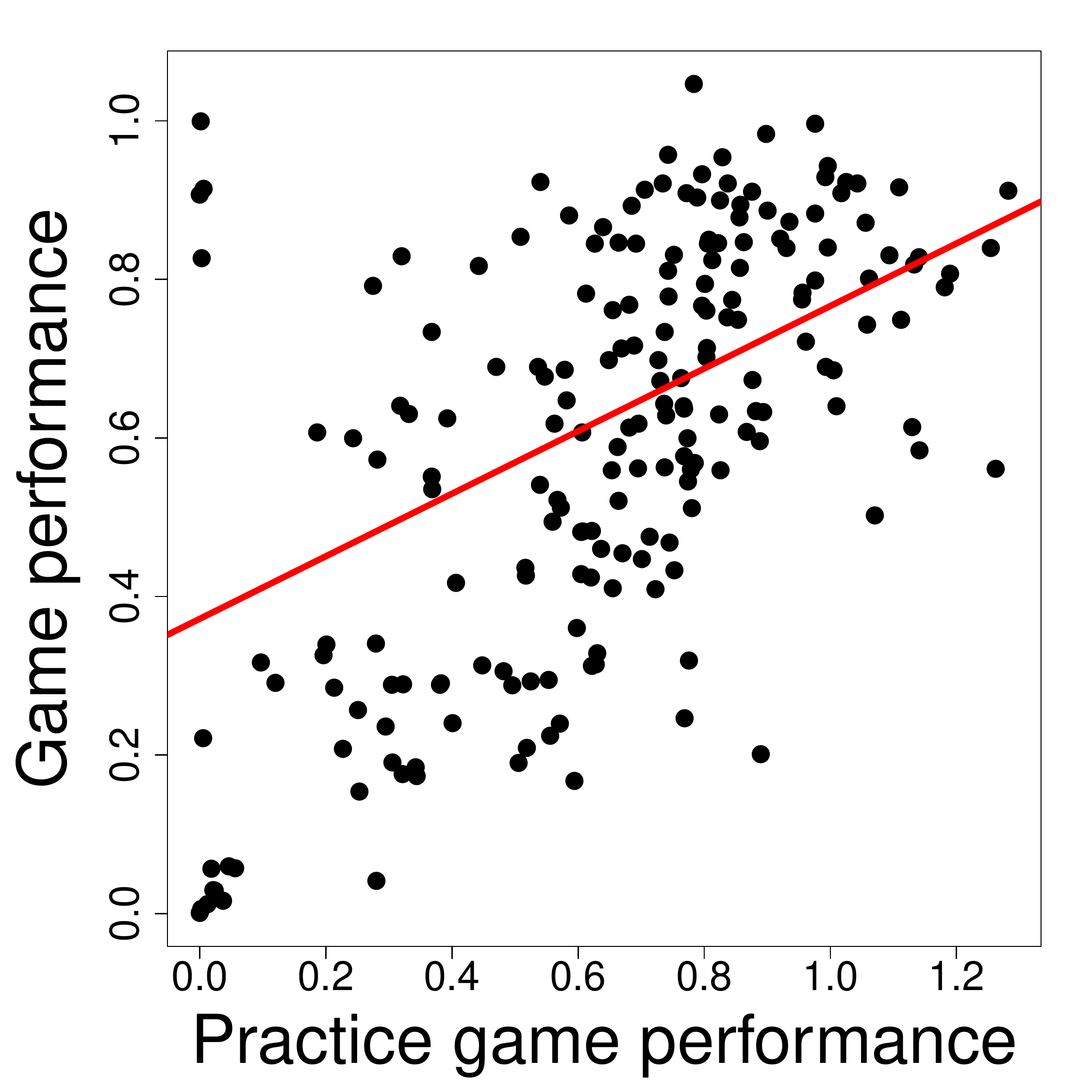}
          \end{minipage} 
          \begin{minipage}{.495\textwidth}
          (b)\\
           \includegraphics[width=\textwidth]{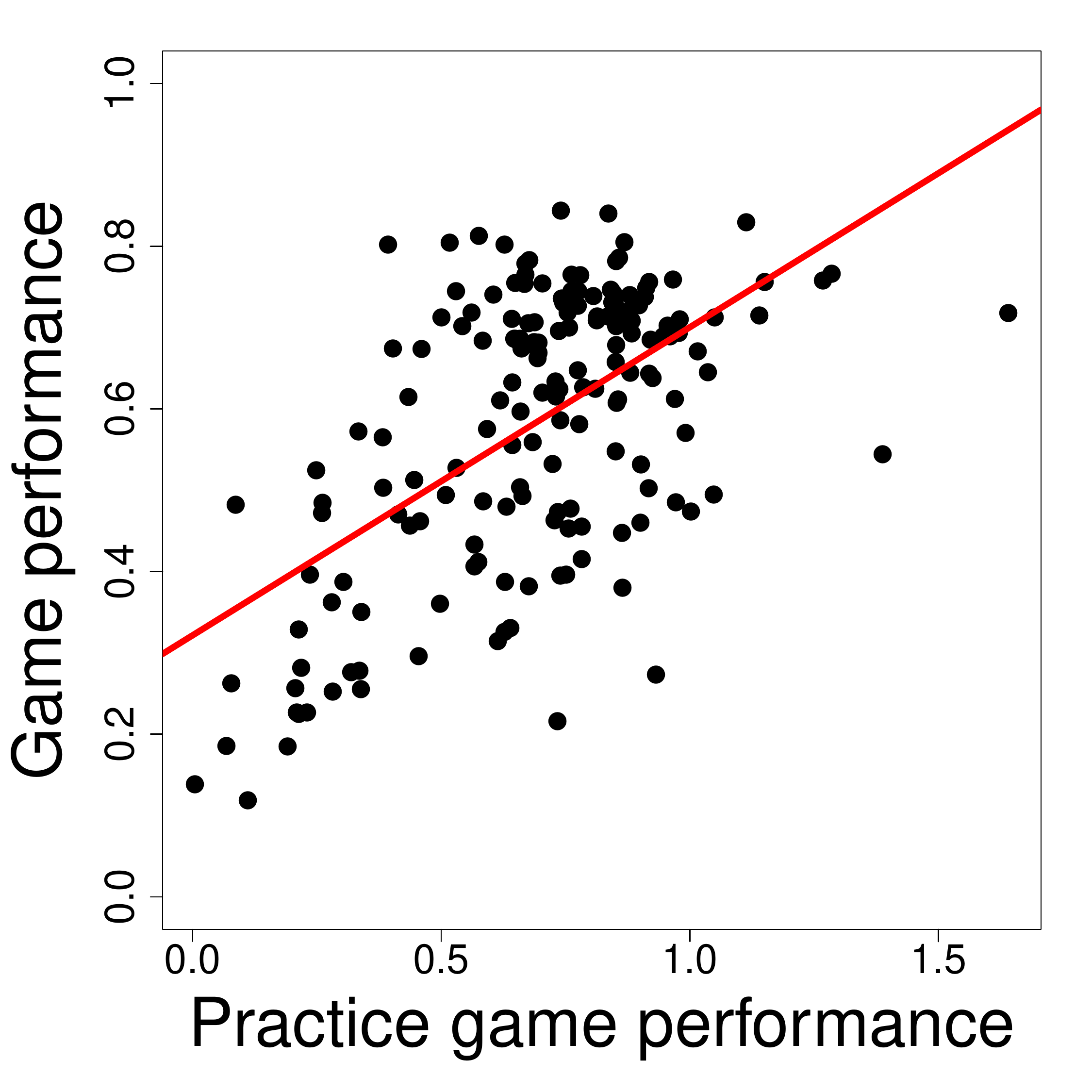}
          \end{minipage}
          \vspace{.1in}
      \caption{Proportion of optimal profit generated by the user in the actual game vs. Proportion of optimal profit generated by the user in the practice, not including the bonus, for (a) the one dimensional game and (b) the age-structured game.}\label{FigPractice}
	\end{figure}

	\begin{figure}[h!]
	  	\includegraphics[width=\textwidth]{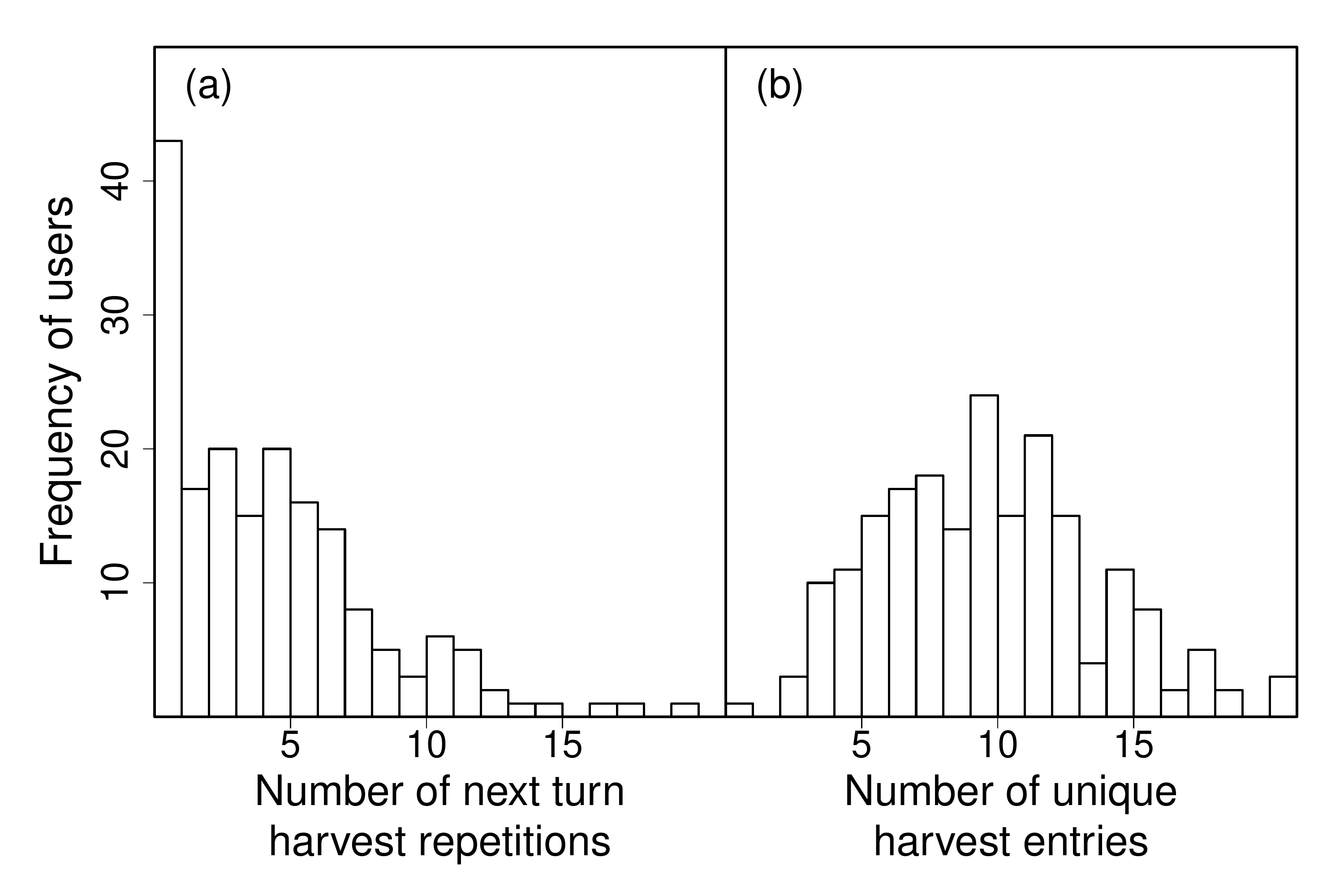}
	  	\vspace{.1in}
	 \caption{(a) A histogram of the number of times the user entered same harvest value, in the one dimensional game, as they did on the previous turn. (b) A histogram of the number of unique harvest entries they made over the course of the entire game.}\label{constHarvesters}
	\end{figure}

  \begin{figure}[h]
      \includegraphics[width=.95\textwidth]{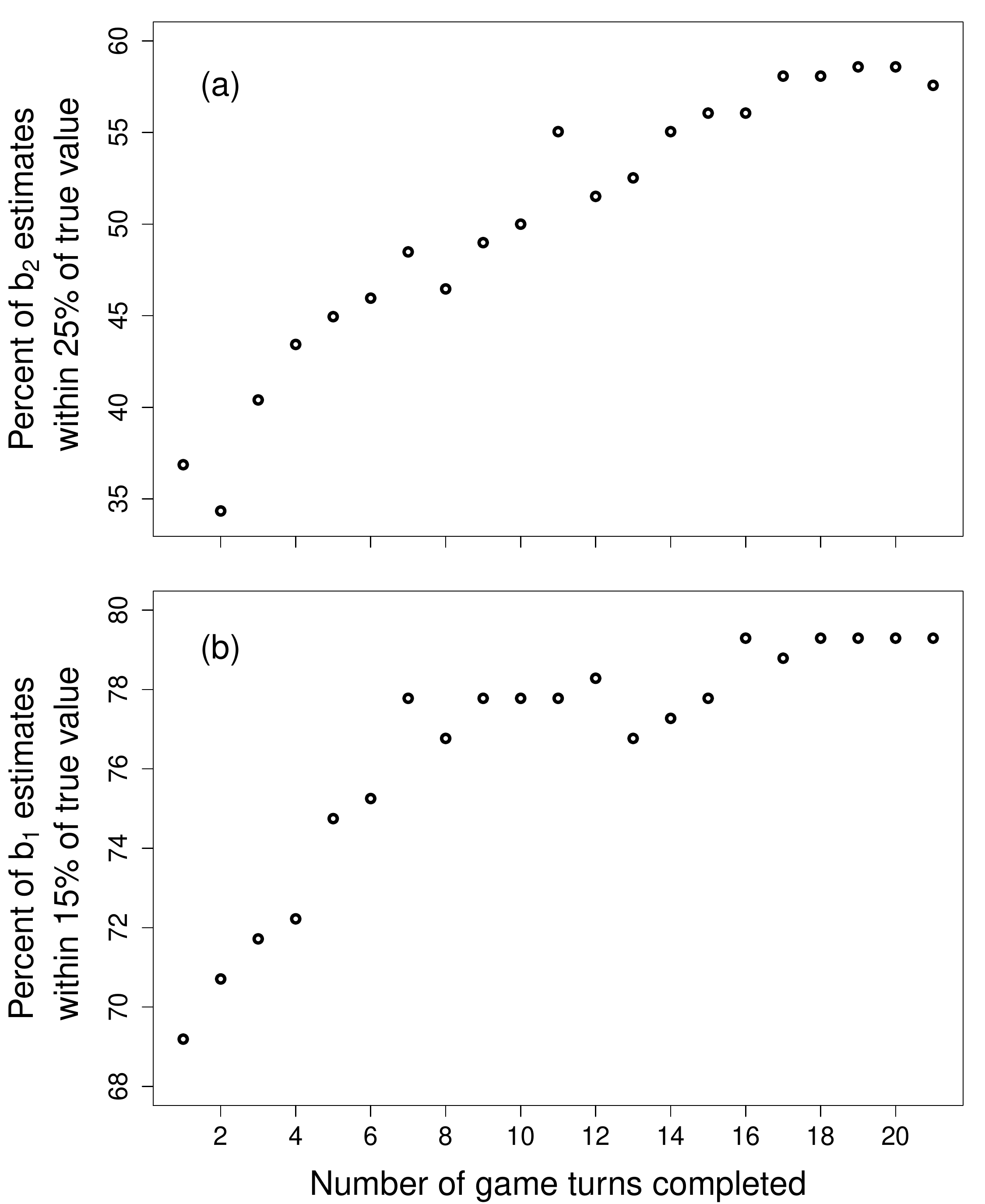}
      \caption{Percent of curve fits where (a) the estimate for $b_2$ was within 25 percent of the true value and (b) the estimate for $b_1$ was within 15 percent of the true value, for each turn during the game.}\label{parmConv}
  \end{figure}

  \begin{figure}[h]
      \includegraphics[width=.98\textwidth]{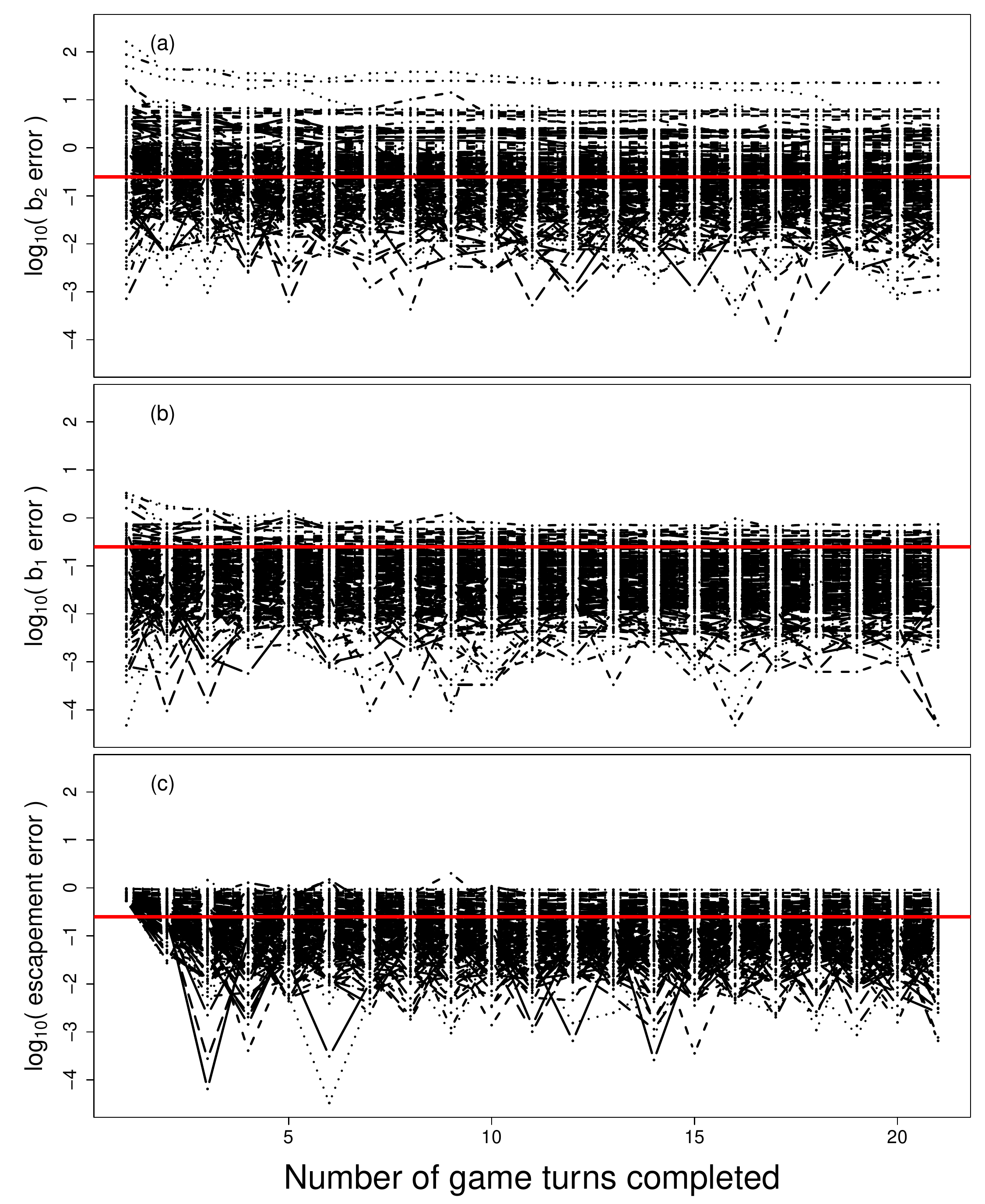}
      \caption{(a) The error in the estimate for $b_1$, $|b_{1,estimate}-b_{1,actual}|/b_{1,actual}$, on log base ten scale when fitting a Beverton-Holt curve through the observed biomass data. A value of zero means the error is the same order of magnitude as the true parameter value. A value of negative one means the estimate is within ten percent of the true value. The red line corresponds to the error being within 25 percent of the true value. (b) The error in the $b_1$ estimate on log 10 scale. (c) The error in the chosen escapement value with the estimates for $b_1$ and $b_2$ relative to optimal escapement.}\label{ParmsFullData}
  \end{figure}


\end{spacing}

\end{document}